\documentclass[]{emulateapj}

\begin{document}

\newcommand{\kms}{km s$^{-1}$}
\newcommand{\msun}{$M_{\odot}$}
\newcommand{\rsun}{$R_{\odot}$}

\title{The ELM Survey. III.  A Successful Targeted Survey for Extremely Low Mass 
White Dwarfs\altaffilmark{*}}

\author{Warren R.\ Brown$^1$,
	Mukremin Kilic$^{2,}$,
	Carlos Allende Prieto$^{3,4}$,
	and Scott J.\ Kenyon$^1$
	}

\affil{ $^1$Smithsonian Astrophysical Observatory, 60 Garden St, Cambridge, MA 02138 USA\\
	$^2$Homer L. Dodge Department of Physics and Astronomy, University of Oklahoma, 440 W. Brooks St., Norman, OK, 73019 USA\\
	$^3$Instituto de Astrof\'{\i}sica de Canarias, E-38205, La Laguna, Tenerife, Spain\\
	$^4$Departamento de Astrof\'{\i}sica, Universidad de La Laguna, E-38206 La Laguna, Tenerife, Spain
	}

\email{wbrown@cfa.harvard.edu, kilic@ou.edu}

\altaffiltext{*}{Based on observations obtained at the MMT Observatory, a joint 
facility of the Smithsonian Institution and the University of Arizona.}

\shorttitle{ Extremely Low Mass White Dwarfs. III. }
\shortauthors{Brown et al.}

\begin{abstract}

	Extremely low mass (ELM) white dwarfs (WDs) with masses $<0.25$ \msun\ are
rare objects that result from compact binary evolution.  Here, we present a targeted
spectroscopic survey of ELM WD candidates selected by color.  The survey is 71\%
complete and has uncovered 18 new ELM WDs.  Of the 7 ELM WDs with follow-up
observations, 6 are short-period binaries and 4 have merger times less than 5 Gyr.  
The most intriguing object, J1741+6526, likely has either a pulsar companion or a
massive WD companion making the system a possible supernova Type Ia or .Ia
progenitor.  The overall ELM Survey has now identified 19 double degenerate binaries
with $<$10 Gyr merger times.  The significant absence of short orbital period ELM
WDs at cool temperatures suggests that common envelope evolution creates ELM WDs
directly in short period systems.  At least one-third of the merging systems are
halo objects, thus ELM WD binaries continue to form and merge in both the disk and
the halo.

\end{abstract}


\slugcomment{Accepted to ApJ}

\section{INTRODUCTION}

	Extremely low-mass (ELM) WDs with masses $<$0.25 \msun\ are created when the
progenitor stars lose so much mass during their evolution that they never
reach helium burning.
	Binary evolution is considered the most likely origin for low-mass WDs
\citep[e.g.][]{marsh95}, making ELM WDs the signposts for the type of compact
systems that are strong gravitational wave sources.
	In Paper I of this series \citep{brown10c} we presented the first complete,
well-defined sample of ELM WDs fortuitously targeted by the Hypervelocity Star
Survey \citep{brown05, brown06, brown09a}.  In paper II of this series
\citep{kilic11a} we characterized other ELM WDs identified in the Sloan Digital Sky
Survey (SDSS) Data Release 4 \citep{eisenstein06}.  In both programs ELM WDs 
exist in $\le1$ day orbital period binary systems, with an estimated merger 
rate comparable to the rate of underluminous supernovae \citep{brown11a}.

	Here we present the results of the first targeted survey for new ELM WDs.  
Our approach is to select high-probability ELM WD candidates from a well-defined
region of color and apparent magnitude, and then obtain spectroscopy for the
objects from that selection region.
	The spectroscopic survey is presently 71\% complete and contains 21 ELM WDs
defined by $5.0<\log{g}<7.0$ dex ($g$ in cm s$^{-2}$).  Three of the ELM WDs were
previously identified \citep{kilic09, brown10c} and 18 are new discoveries.  We have
obtained follow-up spectroscopy for 7 of the new ELM WDs, and calculate orbital
solutions for the 6 with significant velocity variability.  Four of these new ELM WD
systems have gravitational wave merger times less than 5 Gyr.  The most interesting
object, J1741+6526, has a minimum companion mass of 1.1 \msun .  Thus J1741+6526 is
most likely a pulsar binary, or, if the orbit is edge-on, possibly a supernova Type
Ia or .Ia progenitor \citep{bildsten07}.  It will begin mass transfer in $<$170 Myr.

	Our targeted ELM survey will yield a clean, non-kinematically selected
sample of WDs.  Once completed, we can use our sample to constrain the space
density, period distribution, and merger rate of ELM WDs in double degenerate
systems.  SDSS, on the other hand, has not found large numbers of ELM WDs because
they have not targeted them; existing SDSS WD spectroscopy comes from different
target selection programs observed with different completenesses
\citep{eisenstein06}.  In a stellar evolution context, our survey complements
studies of WD binaries with main sequence companions \citep{zorotovic11,
gomezmoran11}. Because of ELM WDs' low surface gravities $\log{g}\sim6$ and small
$\simeq$1 \rsun\ orbital separations, a growing number of systems exhibit some
combination of tidal distortion, relativistic beaming, reflection effects, and
eclipses \citep[e.g.][]{brown11b, kilic11b, parsons11, pyrzas11, steinfadt10,
vennes11}. We expect that on-going photometric follow-up will provide improved
constraints on the nature and orbital inclination of our new ELM WD binary systems.

	We organize this paper as follows.  In Section 2 we discuss our new survey
design, observations, and data analysis.  In Section 3 we present the orbital
solutions for six new ELM WD binaries.  In Section 4 we discuss the overall
properties of the ELM WD sample and highlight correlations between temperature,
orbital period, and secondary mass.  We conclude in Section 5.

\begin{figure}		% FIGURE 1: COLOR-COLOR PLOT
 \plotone{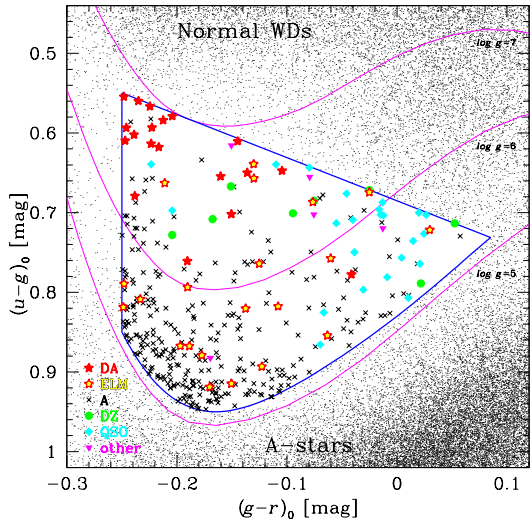}
 \caption{ \label{fig:ugr}
	Color-color diagram showing our ELM Survey selection region (blue bounded
region) in the context of SDSS star counts (black dots).  The ELM Survey selection
targets a region of color space that contains $5<\log{g}<7$ dex hydrogen
atmosphere WDs (magenta solid lines) and avoids known concentrations of A-type stars
and normal WDs.  Symbols indicate the spectroscopic identification of ELM Survey
targets:  red stars are the DA WDs, red stars with yellow centers are the ELM WDs,
black x's are normal A-type stars, green circles are DZ WDs, cyan diamonds are
quasars, and magenta triangles are other miscellaneous objects. }
 \end{figure}

\section{DATA AND ANALYSIS}

\subsection{ELM Survey Design}

	The ELM Survey is a spectroscopic survey of $15<g_0<20$ low mass WD
candidates selected by color.  We use de-reddened, uber-calibrated point spread
function magnitudes from SDSS Data Release 7 \citep[][]{abazajian09}.  Our color
selection strategy, illustrated in Figure \ref{fig:ugr}, is constructed as follows.

	First, we target objects with colors consistent with the effective
temperatures of luminous ELM WDs.  Updated \citet{panei07} evolutionary tracks for
He-core WDs indicate that 0.17 \msun\ WDs spend $\simeq$1 Gyr with luminosities of
$8<M_g<9$ mag at temperatures $\sim$10,000 K.  \citet{serenelli01} tracks give similar
results. Thus we target $-0.25<(g-r)_0<0.1$.
	Second, we target objects with $(u-g)_0$ colors consistent with the
$5<\log{g}<7$ surface gravities of ELM WDs.  To make this color selection, 
we use a polynomial fit to the synthetic photometry of DA WD hydrogen atmosphere
models \citep{koester08},
	\begin{eqnarray}
	(u-g)_0 &<& 0.83 -1.074 (g-r)_0 -1.4939 (g-r)_0^2 \nonumber \\
	& & +0.8156 (g-r)_0^3 +33.42316 (g-r)_0^4 \nonumber \\
	& & +280.88439 (g-r)_0^5 -492.62139 (g-r)_0^6 \nonumber \\
	& & -1993.9254 (g-r)_0^7. \end{eqnarray}
	Finally, we restrict our color-selection to the most probable low mass WD
candidates.  We restrict our selection to those objects bluer in $(u-g)_0$ than the
observed population of A-type stars (our zeropoint in Equation 1) and redder in
$(u-g)_0$ than the observed population of normal DA WDs, $(u-g)_0 > 0.542(g-r)_0 +
0.685$.  We exclude quasars based on their non-stellar color $(r-i)_0 < 1.8 -
0.1g_0$, a limit that becomes more restrictive at faint magnitudes where we expect
greater contamination from increased photometric errors and from increased quasar
number counts.  Put together, this color selection strategy maximizes the
contrast of ELM WDs with respect to foreground and background populations.

	Our target selection identifies 505 ELM WD candidates with $15<g<20$ over
$\simeq$10,000 deg$^2$ of the Sloan Data Release 7 imaging footprint.  Spectra for
116 of these candidates already exist: 31 of the candidates were observed by the
Hypervelocity Star survey \citep{brown09a}, and 85 were observed by SDSS.  Three of
the objects with existing spectra are previously identified merging ELM WD binaries
\citep{kilic09, brown10c}, thus our initial expectation is that we will find about a
dozen new merging ELM WD binaries in the full survey.  There remain 389 candidates
to observe.

\subsection{Spectroscopic Observations}

	We obtained spectra for 245 of the ELM WD candidates in observing runs
starting in 2010 September and ending in 2011 June.  We observed 164 objects with
$g>17$ mag at the 6.5m MMT telescope using the Blue Channel spectrograph
\citep{schmidt89}.  We operated the Blue Channel spectrograph with the 832 line
mm$^{-1}$ grating in second order, providing wavelength coverage 3650 \AA\ to 4500
\AA\ and a spectral resolution of 1.0 - 1.2 \AA , depending on whether a 1\arcsec\
or 1.25\arcsec\ slit was used.  At $g=19$ mag we used a 400 sec exposure time to
obtain a signal-to-noise (S/N) of 7 per pixel in the continuum and a $\simeq$10
\kms\ velocity error.  All objects were observed at the parallactic angle, and a
comparison lamp exposure was obtained with every observation.

	We observed 81 objects with $g<17$ mag in queue scheduled time at the 1.5m
FLWO telescope using the FAST spectrograph \citep{fabricant98}.  We operated FAST
with the 600 line mm$^{-1}$ grating and a 2\arcsec\ slit, providing wavelength
coverage 3500 \AA\ to 5500 \AA\ and a spectral resolution of 2.3 \AA .  At $g=16$
mag we used a 900 sec exposure time to obtain a S/N of 10 per pixel in the
continuum.  A comparison lamp exposure was obtained with every observation.

	We process the spectra using IRAF\footnote[2]{IRAF is distributed by the
National Optical Astronomy Observatories, which are operated by the Association of
Universities for Research in Astronomy, Inc., under cooperative agreement with the
National Science Foundation.} in the standard way.  We flux-calibrate using blue
spectrophotometric standards \citep{massey88}, and we measure radial velocities
using the cross-correlation package RVSAO \citep{kurtz98}.  During the course of the 
ELM Survey we obtained repeat observations for 7 of the newly identified ELM WDs.

\subsection{Spectroscopic Identifications}

	Of the 361 survey targets for which we have spectroscopy:  285 (78.9\%) are
normal A-type stars with $\log{g}\le5$, likely blue horizontal branch stars or blue
stragglers in the halo;  23 (6.4\%) are quasars;  8 (2.2\%) are DZ WDs that show
strong Ca{\sc ii} H and K lines;  5 objects labeled ``other'' in Figure
\ref{fig:ugr} are an emission line galaxy and four featureless continuum objects;
and, most relevant to this paper, 40 objects (11.1\%) are probable DA WDs with
$\log{g}>5$, of which 21 (5.8\% of the survey) are probable ELM WDs with
$5.0<\log{g}<7.0$ dex.

\begin{figure}		% FIGURE 2: SPECTRA
 \plotone{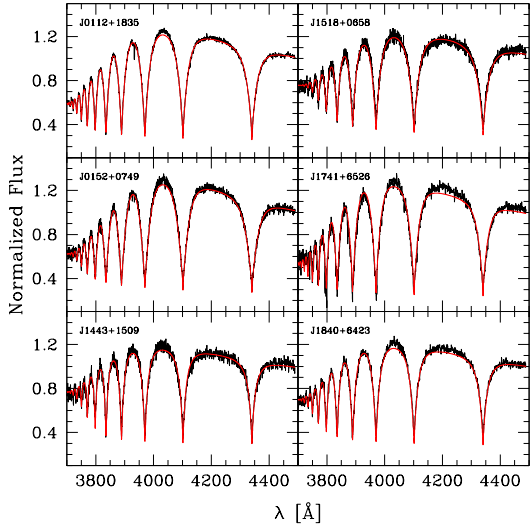}
 \caption{ \label{fig:spec}
	Model fits ({smooth red lines}) overplotted on the composite observed
spectra ({black lines}) for the six newly identified ELM WDs with significant
velocity variability.  We use the spectral continua to provide improved $T_{\rm
eff}$ constraints. }
 \end{figure}

\begin{deluxetable*}{lcccrcrcc}	% TABLE1	- PHYSICAL PARAMETERS -
\tabletypesize{\scriptsize}
\tablecolumns{9}
\tablewidth{0pt}
\tablecaption{DA White Dwarf Physical Parameters\label{tab:param}}
\tablehead{
	\colhead{Object} &
	\colhead{$g_0$} &
	\colhead{$(u-g)_0$} &
	\colhead{$(g-r)_0$}  &
	\colhead{$T_{\rm eff}$} &
	\colhead{$\log g$} &
	\colhead{$M_g$} &
	\colhead{$d_{helio}$} &
	\colhead{Mass} \\
   & (mag) & (mag) & (mag) & (K)~~~~~~ & & (mag)~~~~ & (kpc) & (\msun )
}
	\startdata
J001622.09$-$004323.4 & $19.704 \pm 0.022$ & $0.602 \pm 0.078$ & $-0.239 \pm 0.040$ & $11190 \pm 260$ & $7.40 \pm 0.08$ & $10.90 \pm 0.14$ & 0.58 & 0.36 \\
J011210.25+183503.7   & $17.111 \pm 0.015$ & $0.919 \pm 0.025$ & $-0.170 \pm 0.017$ & $ 9690 \pm 150$ & $5.63 \pm 0.06$ & $ 8.00 \pm 0.20$ & 0.66 & 0.16 \\
J011726.49+251343.2   & $19.300 \pm 0.018$ & $0.679 \pm 0.055$ & $-0.238 \pm 0.027$ & $10710 \pm 190$ & $7.68 \pm 0.08$ & $11.47 \pm 0.15$ & 0.37 & 0.46 \\
J012549.37+461920.1   & $15.835 \pm 0.012$ & $0.778 \pm 0.022$ & $-0.041 \pm 0.014$ & $ 9050 \pm 590$ & $7.18 \pm 0.12$ & $11.30 \pm 0.33$ & 0.08 & 0.29 \\
J015213.77+074913.9   & $18.015 \pm 0.014$ & $0.819 \pm 0.034$ & $-0.249 \pm 0.021$ & $10840 \pm 270$ & $5.80 \pm 0.06$ & $ 7.92 \pm 0.10$ & 1.04 & 0.17 \\
J021847.30+052613.7   & $19.954 \pm 0.027$ & $0.579 \pm 0.069$ & $-0.204 \pm 0.040$ & $13510 \pm 180$ & $7.09 \pm 0.08$ & $ 9.91 \pm 0.18$ & 1.02 & 0.30 \\
J042154.94+830251.7   & $19.789 \pm 0.027$ & $0.618 \pm 0.126$ & $-0.216 \pm 0.039$ & $11870 \pm 200$ & $7.94 \pm 0.07$ & $11.57 \pm 0.11$ & 0.44 & 0.57 \\
J070216.21+111009.0   & $16.061 \pm 0.015$ & $0.764 \pm 0.022$ & $-0.125 \pm 0.019$ & $ 8800 \pm 600$ & $6.00 \pm 0.12$ & $ 9.08 \pm 0.40$ & 0.25 & 0.16 \\
J074511.56+194926.5   & $16.259 \pm 0.008$ & $0.854 \pm 0.018$ & $-0.063 \pm 0.020$ & $ 8190 \pm 550$ & $5.70 \pm 0.12$ & $ 8.00 \pm 0.20$ & 0.45 & 0.16 \\
J074615.83+392203.1   & $16.589 \pm 0.018$ & $0.657 \pm 0.025$ & $-0.130 \pm 0.026$ & $12130 \pm 400$ & $5.98 \pm 0.12$ & $ 8.44 \pm 0.39$ & 0.43 & 0.17 \\
J082904.78+370518.4   & $19.242 \pm 0.017$ & $0.593 \pm 0.048$ & $-0.246 \pm 0.026$ & $13430 \pm 230$ & $7.42 \pm 0.08$ & $10.48 \pm 0.14$ & 0.57 & 0.38 \\
J084325.09+371551.7   & $19.890 \pm 0.034$ & $0.702 \pm 0.081$ & $-0.150 \pm 0.044$ & $12950 \pm 280$ & $7.79 \pm 0.08$ & $11.19 \pm 0.17$ & 0.55 & 0.50 \\
J090052.04+023413.8   & $17.965 \pm 0.023$ & $0.639 \pm 0.033$ & $-0.130 \pm 0.029$ & $ 8220 \pm 330$ & $5.78 \pm 0.07$ & $ 8.00 \pm 0.20$ & 0.98 & 0.16 \\
J091826.05+375308.7   & $18.607 \pm 0.017$ & $0.560 \pm 0.032$ & $-0.235 \pm 0.034$ & $12060 \pm 250$ & $7.91 \pm 0.08$ & $11.53 \pm 0.14$ & 0.26 & 0.56 \\
J095353.66+410927.4   & $19.589 \pm 0.022$ & $0.554 \pm 0.053$ & $-0.248 \pm 0.030$ & $13040 \pm 490$ & $7.65 \pm 0.08$ & $10.91 \pm 0.16$ & 0.54 & 0.46 \\
J111215.82+111745.0   & $16.235 \pm 0.017$ & $0.820 \pm 0.034$ & $-0.138 \pm 0.026$ & $ 9400 \pm 490$ & $5.81 \pm 0.12$ & $ 8.38 \pm 0.27$ & 0.37 & 0.16 \\
J113017.45+385550.1   & $19.413 \pm 0.029$ & $0.761 \pm 0.062$ & $-0.191 \pm 0.045$ & $12030 \pm 180$ & $7.17 \pm 0.08$ & $10.32 \pm 0.12$ & 0.66 & 0.30 \\
J113723.44+123105.9   & $19.122 \pm 0.031$ & $0.654 \pm 0.059$ & $-0.160 \pm 0.036$ & $13000 \pm 390$ & $7.79 \pm 0.08$ & $11.19 \pm 0.18$ & 0.39 & 0.50 \\
J114303.83+361843.8   & $19.678 \pm 0.021$ & $0.567 \pm 0.057$ & $-0.225 \pm 0.035$ & $15090 \pm 190$ & $7.62 \pm 0.08$ & $10.54 \pm 0.15$ & 0.67 & 0.46 \\
J123316.19+160204.6\tablenotemark{a}   & $19.829 \pm 0.018$ & $0.809 \pm 0.068$ & $-0.233 \pm 0.028$ & $10920 \pm 190$ & $5.12 \pm 0.07$ & $ 8.00 \pm 0.20$ & 2.32 & 0.17 \\
J123523.78+475029.1   & $19.291 \pm 0.018$ & $0.650 \pm 0.080$ & $-0.136 \pm 0.024$ & $10600 \pm 200$ & $7.68 \pm 0.08$ & $11.45 \pm 0.13$ & 0.37 & 0.46 \\
J144342.74+150938.6   & $18.578 \pm 0.017$ & $0.757 \pm 0.031$ & $-0.060 \pm 0.023$ & $ 8810 \pm 220$ & $6.32 \pm 0.07$ & $ 9.67 \pm 0.23$ & 0.61 & 0.17 \\
J151826.68+065813.2   & $17.456 \pm 0.019$ & $0.687 \pm 0.026$ & $-0.076 \pm 0.024$ & $ 9810 \pm 320$ & $6.66 \pm 0.06$ & $10.08 \pm 0.18$ & 0.30 & 0.20 \\
J152122.59+032607.1   & $18.439 \pm 0.017$ & $0.794 \pm 0.041$ & $-0.190 \pm 0.022$ & $15210 \pm 250$ & $5.60 \pm 0.08$ & $ 8.00 \pm 0.20$ & 1.22 & 0.17 \\
J152651.57+054335.3   & $18.735 \pm 0.018$ & $0.789 \pm 0.041$ & $-0.248 \pm 0.028$ & $10340 \pm 200$ & $5.64 \pm 0.08$ & $ 8.00 \pm 0.20$ & 1.40 & 0.17 \\
J153300.03+492948.3   & $19.221 \pm 0.021$ & $0.610 \pm 0.047$ & $-0.247 \pm 0.030$ & $15830 \pm 320$ & $7.75 \pm 0.08$ & $10.68 \pm 0.20$ & 0.51 & 0.49 \\
J161431.28+191219.4   & $16.187 \pm 0.019$ & $0.893 \pm 0.025$ & $-0.123 \pm 0.024$ & $ 8590 \pm 540$ & $5.64 \pm 0.12$ & $ 8.00 \pm 0.20$ & 0.43 & 0.16 \\
J161722.51+131018.8   & $18.605 \pm 0.014$ & $0.868 \pm 0.041$ & $-0.189 \pm 0.021$ & $10480 \pm 240$ & $5.78 \pm 0.08$ & $ 8.00 \pm 0.20$ & 1.32 & 0.17 \\
J174140.49+652638.7   & $18.271 \pm 0.022$ & $0.879 \pm 0.047$ & $-0.177 \pm 0.030$ & $ 9790 \pm 240$ & $5.19 \pm 0.06$ & $ 8.00 \pm 0.20$ & 1.13 & 0.16 \\
J184037.78+642312.3   & $18.757 \pm 0.014$ & $0.818 \pm 0.047$ & $-0.108 \pm 0.019$ & $ 9140 \pm 170$ & $6.16 \pm 0.06$ & $ 9.25 \pm 0.13$ & 0.80 & 0.17 \\
J191311.59+372631.7   & $15.430 \pm 0.007$ & $0.722 \pm 0.016$ & $+0.030 \pm 0.008$ & $ 8070 \pm 630$ & $7.04 \pm 0.12$ & $11.55 \pm 0.40$ & 0.06 & 0.25 \\
J211921.96$-$001825.8\tablenotemark{a}   & $20.000 \pm 0.021$ & $0.867 \pm 0.092$ & $-0.197 \pm 0.033$ & $10360 \pm 250$ & $5.36 \pm 0.07$ & $ 8.00 \pm 0.20$ & 2.51 & 0.17 \\
J213513.09$-$072442.5 & $19.533 \pm 0.020$ & $0.584 \pm 0.060$ & $-0.212 \pm 0.030$ & $11230 \pm 260$ & $7.30 \pm 0.08$ & $10.70 \pm 0.16$ & 0.58 & 0.33 \\
J221928.48+120418.6   & $17.681 \pm 0.017$ & $0.915 \pm 0.031$ & $-0.150 \pm 0.022$ & $ 9700 \pm 160$ & $5.12 \pm 0.08$ & $ 8.00 \pm 0.20$ & 0.86 & 0.16 \\
J221936.32$-$092617.4 & $19.887 \pm 0.036$ & $0.593 \pm 0.126$ & $-0.223 \pm 0.053$ & $14490 \pm 250$ & $7.83 \pm 0.08$ & $11.03 \pm 0.12$ & 0.59 & 0.51 \\
J222859.93+362359.6   & $16.522 \pm 0.011$ & $0.674 \pm 0.019$ & $-0.025 \pm 0.014$ & $ 8840 \pm 540$ & $6.73 \pm 0.12$ & $10.63 \pm 0.37$ & 0.15 & 0.20 \\
J223630.08+223223.8\tablenotemark{b}   & $17.004 \pm 0.021$ & $0.663 \pm 0.028$ & $-0.211 \pm 0.026$ & $11290 \pm 50$ & $6.30 \pm 0.02$ & $ 8.83 \pm 0.27$ & 0.43 & 0.17 \\
J231757.41+060252.1   & $19.449 \pm 0.035$ & $0.647 \pm 0.065$ & $-0.104 \pm 0.040$ & $12010 \pm 280$ & $6.98 \pm 0.08$ & $10.03 \pm 0.16$ & 0.76 & 0.26 \\
J231910.03+175824.6   & $19.448 \pm 0.017$ & $0.610 \pm 0.052$ & $-0.145 \pm 0.024$ & $13840 \pm 190$ & $7.19 \pm 0.08$ & $10.02 \pm 0.14$ & 0.77 & 0.32 \\
J233705.02+153353.6   & $19.676 \pm 0.022$ & $0.614 \pm 0.063$ & $-0.224 \pm 0.033$ & $11700 \pm 180$ & $7.63 \pm 0.07$ & $11.11 \pm 0.11$ & 0.52 & 0.45 \\
	\enddata
\tablenotetext{a}{\citet{brown10c}}
\tablenotetext{b}{\citet{kilic09}}
\end{deluxetable*}

\subsection{Stellar Atmosphere Parameters}

	We perform stellar atmosphere model fits using an upgraded version of the 
code described by \citet{allendeprieto06} and synthetic DA WD spectra kindly 
provided by D.\ Koester.  The grid of WD model atmospheres covers effective 
temperatures from 6000 K to 30,000 K in steps of 500 K to 2000 K, and surface 
gravities from $\log{g}=$ 5.0 to 9.0 in steps of 0.25 dex.  The model atmospheres 
are calculated assuming local thermodynamic equilibrium and include both convective 
and radiative transport \citep{koester08}.

\begin{figure}		% FIGURE 3: TEFF - LOGG PLOT
 \plotone{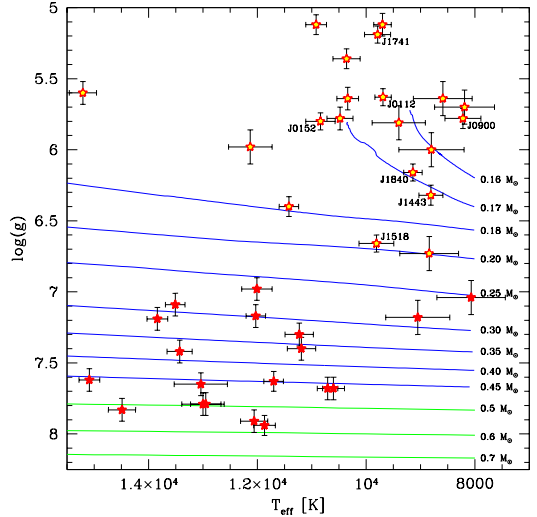}
 \caption{ \label{fig:teff}
	Surface gravity vs.\ effective temperature of the observed DA WDs (solid red
stars) and ELM WDs with $<$0.25 \msun\ (stars with yellow centers) found in this
survey, compared with predicted tracks for He WDs with 0.16--0.45 \msun\ \citep[{\it
blue lines},][]{panei07} and CO WDs with 0.5--0.7 \msun\ \citep[{\it green
lines},][]{bergeron95,holberg06}.}
 \end{figure}

	We fit the full flux-calibrated spectra as well as the continuum-corrected
Balmer line profiles.  The spectral continuum provides improved constraints on
effective temperature but exposes our fits to possible flux calibration problems --
a concern for some of the objects observed in the 1.5m FAST queue, but generally not
an issue for the $g>17$ mag objects observed at the MMT.  When we compare best-fit
solutions, the flux-calibrated and continuum-corrected parameters differ on average
by $540 \pm 700$ K in $T_{\rm eff}$ and $0.06 \pm 0.08$ dex in $\log g$.  We take
these differences as our systematic error.  We consider the flux-calibrated
parameters more robust, because synthetic photometry of the flux-calibrated model
fits provide better agreement with SDSS photometry in all five filters.

% $1316\pm400$ and $153\pm618$ K in $T_{\rm eff}$, and $0.29\pm0.21$ and $0.01\pm0.18$ in $\log g$.

	Table \ref{tab:param} presents the $T_{\rm eff}$ and $\log{g}$ values for
the 40 DA WDs identified in the ELM Survey.  For the 7 newly identified ELM WDs with
multiple observations we use the flux-calibrated, summed spectra to derive $T_{\rm
eff}$ and $\log{g}$, and we use the scatter of the fits to the individual spectra to
derive errors.  The remaining DA WDs typically have single-epoch, S/N$\simeq$8 per
pixel spectra and increased statistical uncertainties in their atmospheric
parameters.  Two of the more massive DA WDs, J084325.09+371551.7 and
J095353.66+410927.4, have SDSS stellar atmosphere measurements \citep{eisenstein06}
that differ from our measurements at the 1- to 2-$\sigma$ level.  Figure
\ref{fig:spec} visually compares our best-fit atmosphere models with the summed
spectra of the six ELM WDs with radial velocity variability.  We attribute imperfect
continuua fits to imperfect flux calibration, notably for J1741+6528 and J1840+6423
which were observed at high airmass.

	Figure \ref{fig:teff} plots the $T_{\rm eff}$ and $\log g$ of all 40 DA
WDs in our targeted survey in relation to the improved \citet{panei07} tracks
\citep[see][]{kilic10} for He-core WDs with masses 0.16--0.45 \msun\ and the
\citet{bergeron95}\footnote[3]{http://www.astro.umontreal.ca/$\sim$bergeron/CoolingModels/}
tracks for normal CO-core WDs with masses 0.5--0.7 \msun . The gap between the 0.17
\msun\ and 0.18 \msun\ He-core WD tracks is linked to the threshold for
thermonuclear flashes in the hydrogen shell burning phase \citep{panei07}.
	The presence of objects with $\log{g}<6$ in the gap between tracks makes
precise WD mass and luminosity estimates difficult \citep[see][]{kilic11a,vennes11}.  
More reliable estimates are possible for the $\log{g}>6$ WDs.  Mass, luminosity, and
heliocentric distance estimates are presented in Table \ref{tab:param}.

\subsection{Radial Velocities}

	We maximize our sensitivity to velocity variability with the following
approach.  We first cross-correlate the individual spectra with a high
signal-to-noise WD template.  We then shift the individual spectra to the rest
frame, and sum them together to create a template for each object.  Finally, we
cross-correlate the individual spectra with the appropriate template to obtain the
final velocities for each object.  The average precision of our measurements is
$\pm12$ \kms .  We verify our velocities using WD model spectra with atmospheric
parameters customized for each target. The results are consistent within 10 km
s$^{-1}$, which we take as our systematic velocity uncertainty.
	The Appendix data table presents the full set of radial velocity
measurements for the 7 newly identified ELM WDs with multiple observations.

	Six of the 7 newly identified ELM WDs with multiple observations display
significant radial velocity variability.
	The ELM WD with no significant radial velocity variability is J0900+0234,
however we cannot rule out whether it is a binary or not.  We first observed
J0900+0234 with a single exposure on 2011 March 3 and it had a heliocentric radial
velocity of $64\pm12$ \kms .  On 2011 May 9 we re-observed the ELM WD with 9
back-to-back 2.5 min exposures and it had a mean 67 \kms\ velocity with a $\pm24$
\kms\ dispersion.  Although the measurement error was $\pm13$ \kms , half of the
observed dispersion, there was no obvious periodicity.  Thus J0900+0234 shows no
significant velocity variation over the observed time baselines.  A future epoch of
observations is required to rule out the possibility that our existing observations
may have sampled the same orbital phase.  Follow-up observations are planned for the
other ELM WDs as well.

\begin{figure}		% FIGURE 4: PERIODOGRAM
 % \plotone{/pool/wbrown0/Bcand/LMWD3/pdm.ps}
 \plotone{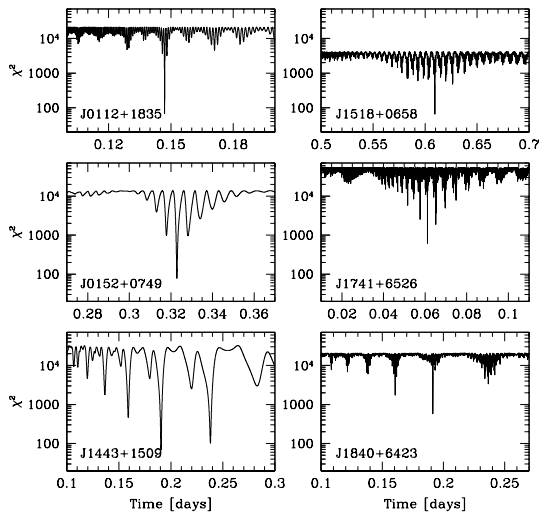}
 \caption{ \label{fig:pdm}
	Periodograms for the 6 newly identified ELM WDs with significant velocity
variability.  Some objects are well constrained, while others have multiple period
aliases. In all cases the periods are $<$1 day. }
 \end{figure}

%TABLE2
\begin{deluxetable*}{lcrrcccc}
\tabletypesize{\scriptsize}
\tablecolumns{8}
\tablewidth{0pt}
\tablecaption{Binary Orbital Parameters\label{tab:orbit}}
\tablehead{
\colhead{Object}&
\colhead{$P$}&
\colhead{$K$}&
\colhead{$\gamma$}&
\colhead{Spec.\ Conjunction}&
\colhead{Mass Function}&
\colhead{$M_2$}&
\colhead{$\tau_{\rm merge}$}\\
  & (days) & (km s$^{-1}$) & (km s$^{-1}$) & (days + 2455250) & & (\msun ) & (Gyr)
}
	\startdata
J011210.25+183503.7    & $0.14698 \pm 0.00003$  & $295 \pm  2$ & $-121 \pm  1$ &  $ 259.61940 \pm 0.00015$ & $0.392 \pm 0.007$  & $\ge$ 0.62 & $\le$ 2.7 \\
J015213.77+074913.9    & $0.32288 \pm 0.00014$  & $217 \pm  2$ & $ -61 \pm  1$ &  $ 261.50244 \pm 0.00040$ & $0.341 \pm 0.008$  & $\ge$ 0.57 & $\le$ 22  \\
J090052.04+023413.8\tablenotemark{a} & \nodata  &   $\le 24$   & $  67 \pm  8$ &           \nodata         &       \nodata      &   \nodata  & \nodata   \\
J144342.74+150938.6    & $0.19053 \pm 0.02402$  & $307 \pm  3$ & $-172 \pm  3$ &  $ 490.77780 \pm 0.00057$ & $0.569 \pm 0.074$  & $\ge$ 0.83 & $\le$ 4.1 \\
J151826.68+065813.2    & $0.60935 \pm 0.00004$  & $172 \pm  2$ & $ -67 \pm  2$ &  $  25.98375 \pm 0.00131$ & $0.322 \pm 0.005$  & $\ge$ 0.58 & $\le$ 101 \\
J174140.49+652638.7    & $0.06111 \pm 0.00001$  & $508 \pm  4$ & $ -70 \pm  3$ &  $ 259.54392 \pm 0.00006$ & $0.830 \pm 0.018$  & $\ge$ 1.10 & $\le$ 0.17\\
J184037.78+642312.3    & $0.19130 \pm 0.00005$  & $272 \pm  2$ & $ -76 \pm  2$ &  $ 259.43724 \pm 0.00086$ & $0.399 \pm 0.009$  & $\ge$ 0.64 & $\le$ 5.0 \\
	\enddata
\tablenotetext{a}{Existing measurements are consistent with no variation.}
\end{deluxetable*}

\begin{figure*}		% FIGURE 5: VELOCITY PLOTS
 \centerline{
	\includegraphics[height=2.5in]{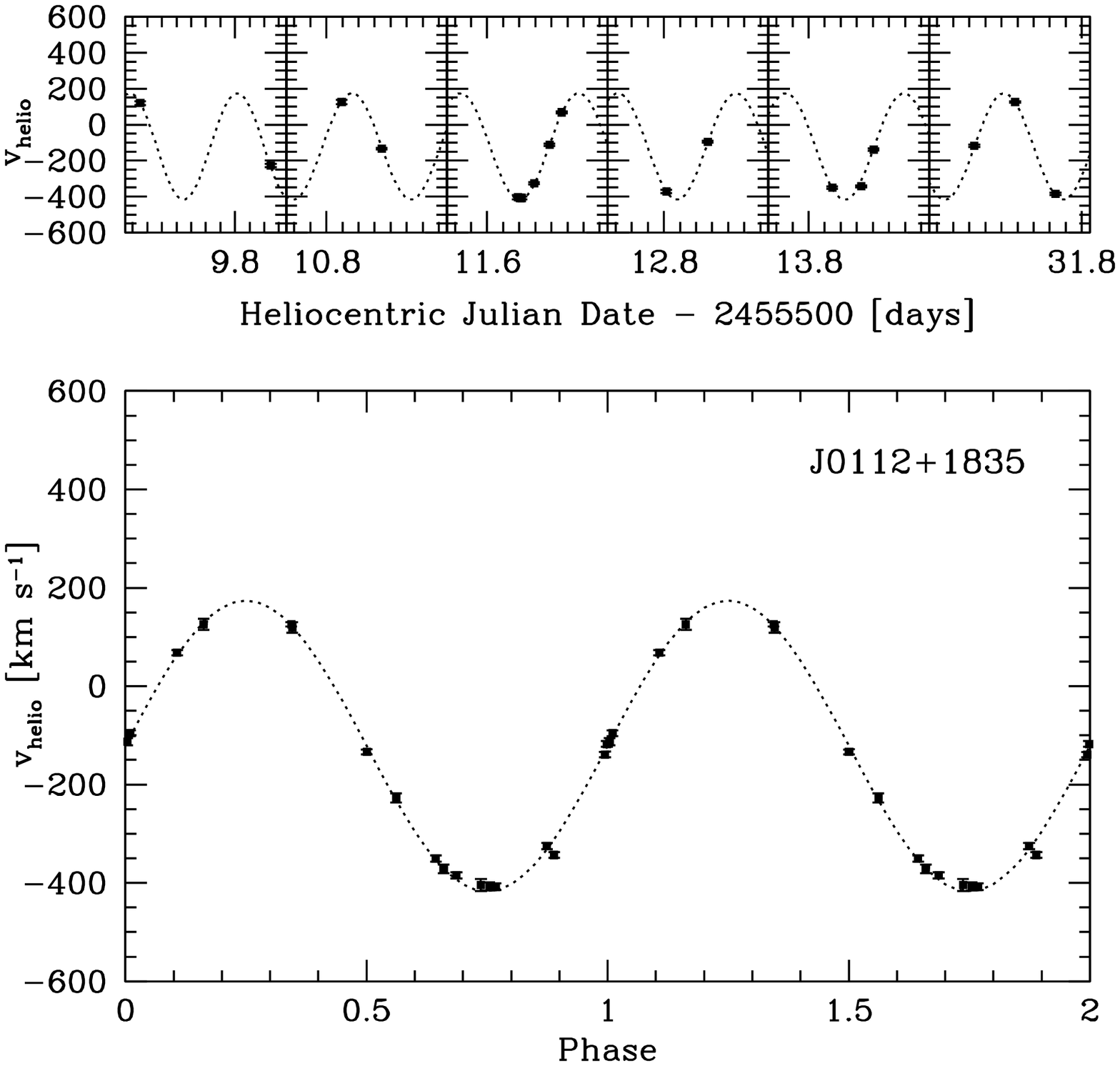} ~~~~~
	\includegraphics[height=2.5in]{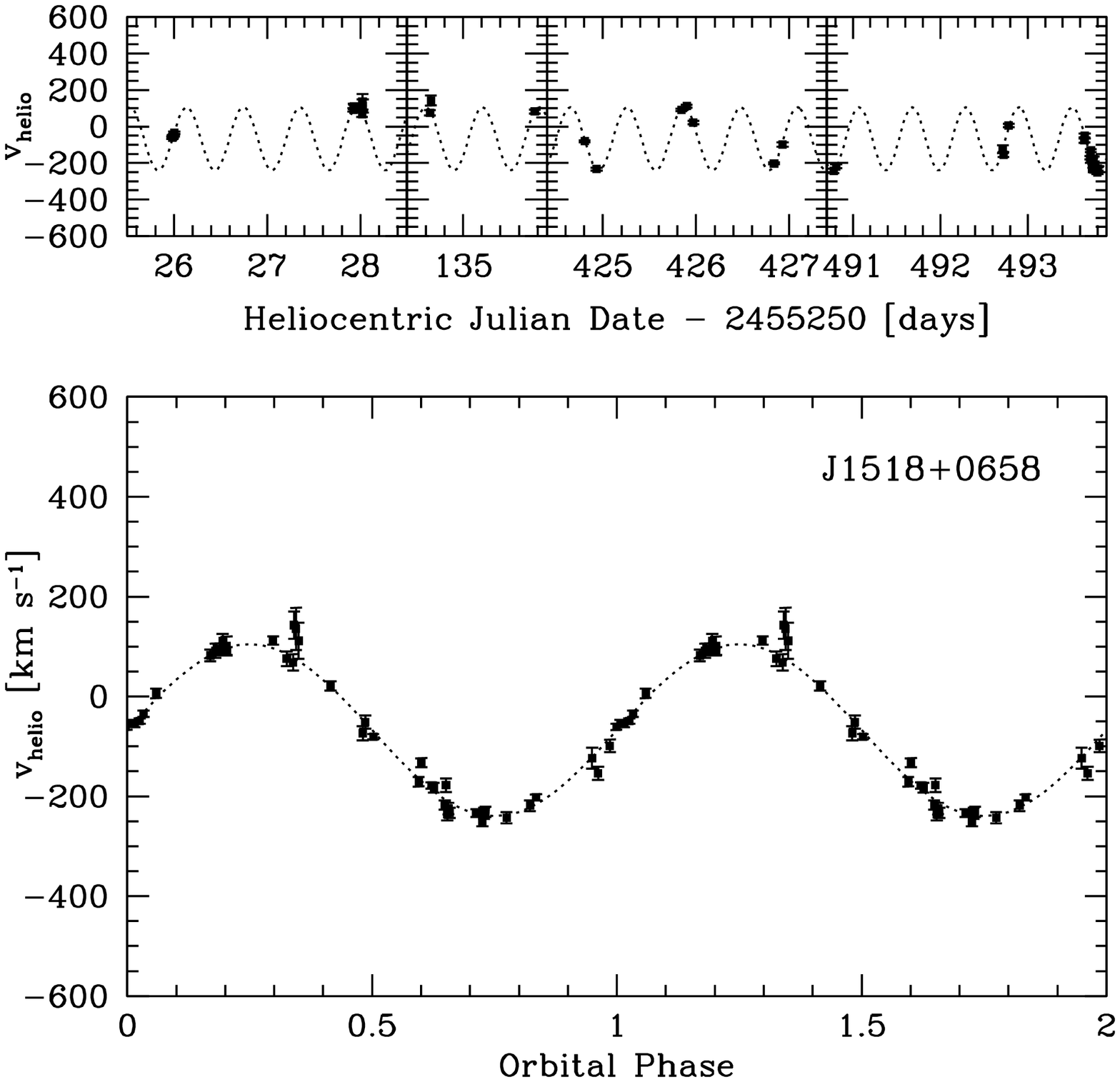} }
 \centerline{
	\includegraphics[height=2.5in]{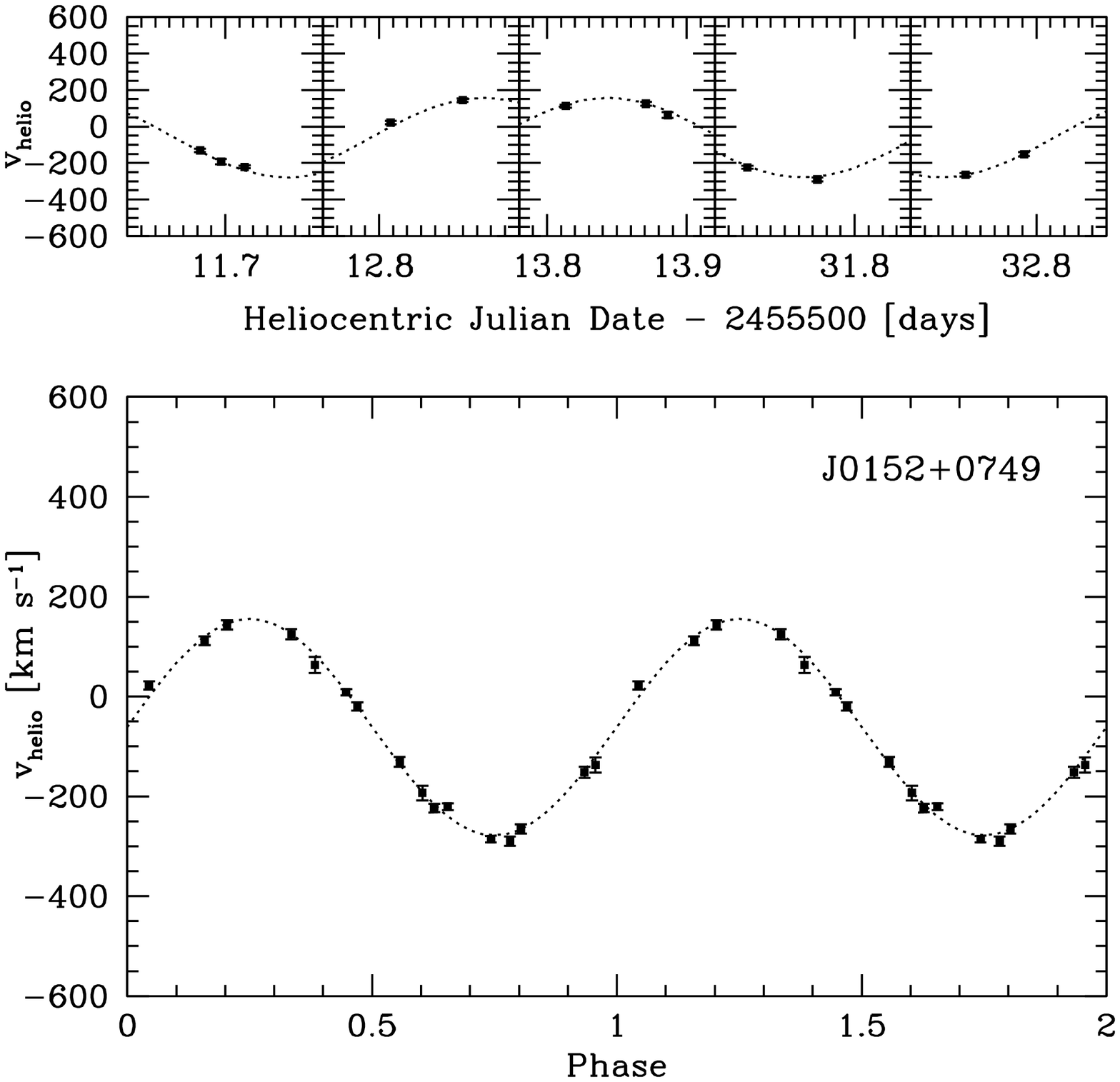} ~~~~~
	\includegraphics[height=2.5in]{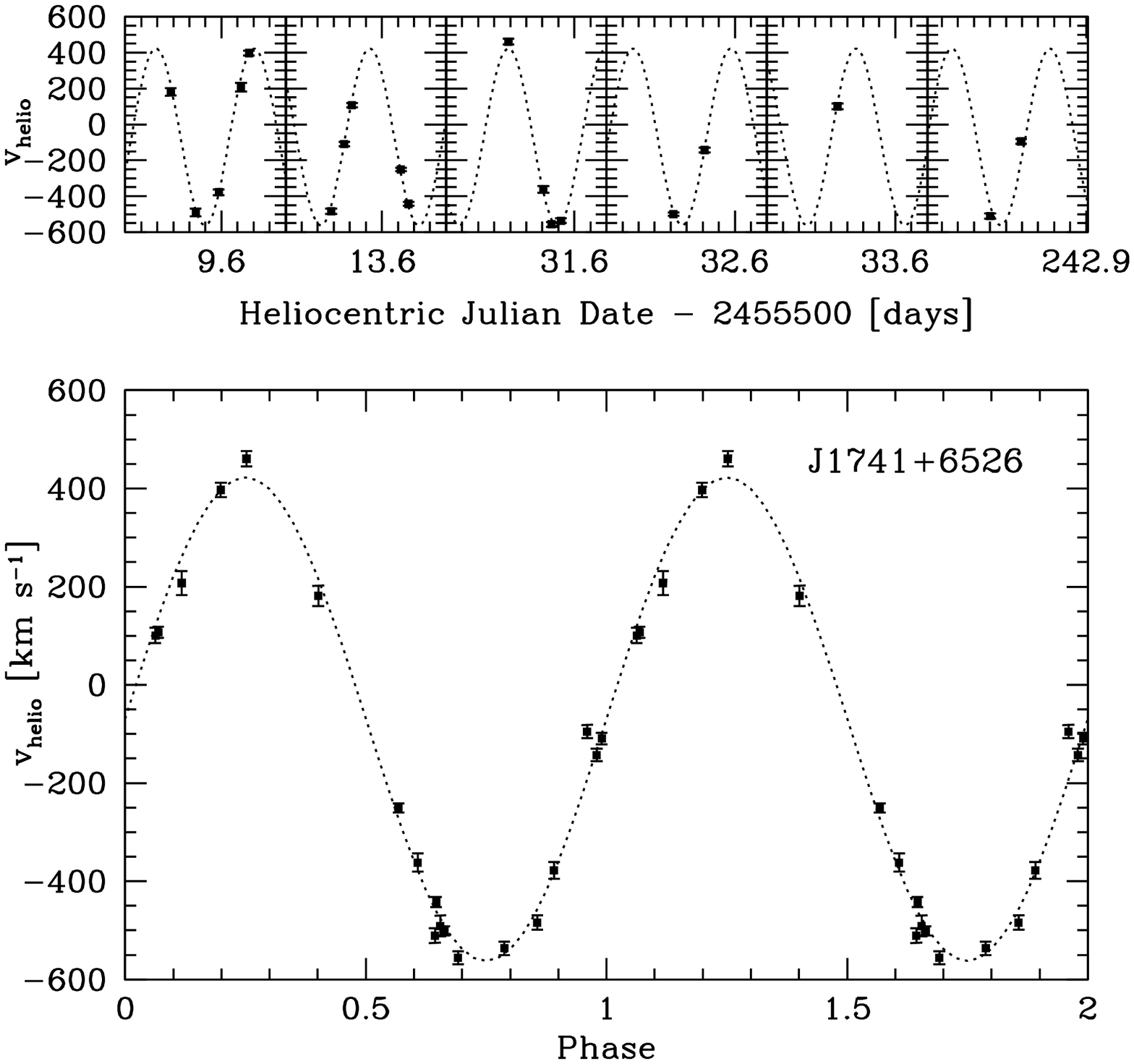} }
 \centerline{
	\includegraphics[height=2.5in]{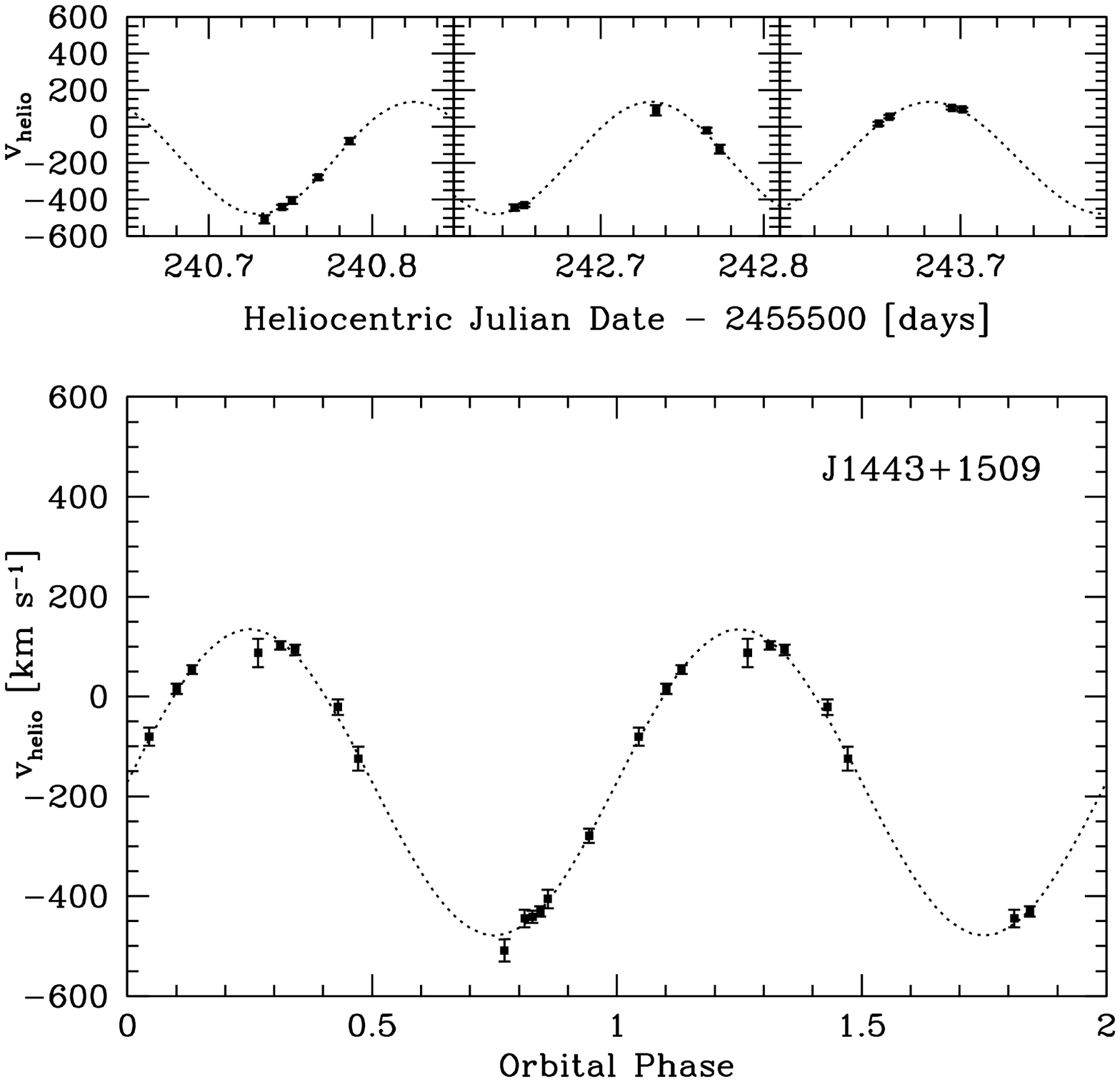} ~~~~~
	\includegraphics[height=2.5in]{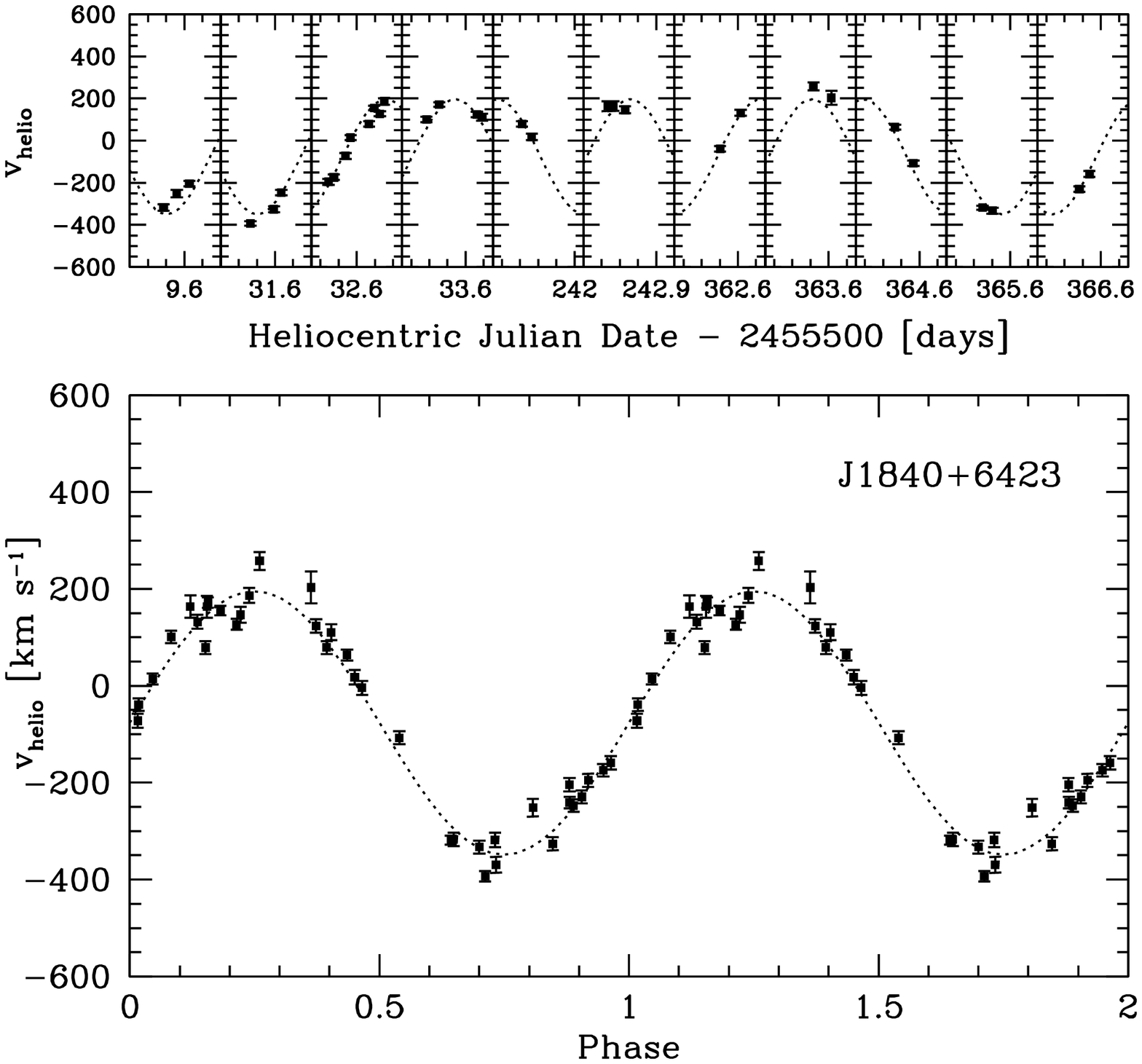} }
 \caption{ \label{fig:vel}
	Observed velocities and best-fit orbits for the 6 newly identified ELM WD
binaries.  Small panels plot the heliocentric radial velocities vs.\ observation
date.  Large panels plot the observations phased to the best-fit orbit (Table
\ref{tab:orbit}).  The same vertical axis is used in all panels. }
 \end{figure*}

\subsection{Orbital Elements}

	We now compute the orbital period and other orbital elements for the six
ELM WDs with significant radial velocity variability.  We begin by solving for the
best-fit period that minimizes $\chi^2$ for a circular orbit.  Figure \ref{fig:pdm}
plots the periodograms.  In some cases there are multiple period solutions because
of insufficient coverage, however in all cases the periods are constrained to be
$<$1 day.  We estimate the period error by conservatively identifying the range of
periods with $\chi^2 \le 2 \chi^2_{\rm min}$, where $\chi^2_{\rm min}$ is the
minimum $\chi^2$.

	We compute best-fit orbital elements using the code of \citet{kenyon86},
which weights each velocity measurement by its associated error.  The uncertainties
in the orbital elements are derived from the covariance matrix and $\chi^2$.  To
verify these uncertainty estimates, we perform a Monte Carlo analysis where we
replace the measured radial velocity $v$ with $v + g \delta v$, where $\delta v$ is
the error in $v$ and $g$ is a Gaussian deviate with zero mean and unit variance.  
For each of 10000 sets of modified radial velocities, we repeat the periodogram
analysis and derive new orbital elements. We adopt the inter-quartile range in the
period and orbital elements as the uncertainty.  For binaries with multiple period
aliases, both approaches yield similar uncertainties.  When there are several
equally plausible periods, the Monte Carlo analysis selects all possible periods and
derives very large uncertainties.  In these cases, we adopt errors from the
covariance matrix for the lowest $\chi^2$ orbital period.  We plot the best-fit
orbits in Figure \ref{fig:vel}.

	Table \ref{tab:orbit} presents the best-fit orbital parameters. Columns
include orbital period ($P$), radial velocity semi-amplitude ($K$), systemic
velocity ($\gamma$), the time of spectroscopic conjunction (the time when the object
is closest to us), mass function (see Eqn.\ 1 below), and minimum secondary mass
(assuming $i=90\arcdeg$).  The systemic velocities in Table \ref{tab:orbit} are not
corrected for the WDs' gravitational redshifts, which should be subtracted from the
observed velocities to find the true systemic velocities. This correction is a few
km s$^{-1}$ for a 0.17 \msun\ helium WD, comparable to the systemic velocity
uncertainty.

\section{RESULTS}

	The orbital solutions constrain the mass and thus the nature of the ELM WD
binary companions, as well as the binary systems' gravitational wave merger times.  
We discuss each binary in turn.

\subsection{J011210.25+183503.7}

	The ELM WD J0112+1835 has a well-constrained orbital period of 
$3.5275\pm0.0006$ hr and a radial velocity amplitude of $590\pm4$ \kms .
Its binary mass function is given by
	\begin{equation}
\frac{M_2^3~{\rm sin}^3i}{(M_1+M_2)^2}=\frac{P K^3}{2 \pi G}= 0.392 \pm 0.007 M_{\sun},
	\end{equation} where $i$ is the orbital inclination angle, $M_1\simeq0.16$
\msun\ is the ELM WD mass inferred from \citet{panei07} tracks, and $M_2$ is the
companion mass.  For an edge-on orbit with $i=90\arcdeg$, Eqn.\ 2 provides the
minimum companion mass (see Table \ref{tab:orbit}).  Assuming a random orbital
inclination distribution, on the other hand, allows us to calculate the probability
of different companion masses.

	Given the observed orbital parameters, there is a 71\% probability that
J0112+1835's unseen companion is a WD with $<$1.4 \msun\ and a 14\% probability that
the companion is a neutron star with 1.4-3.0 \msun . The likelihood that the system
contains a pair of WDs whose total mass exceeds the Chandrasekhar mass is 4\%. If we
assume the mean inclination angle for a random stellar sample, $i=60\arcdeg$, we get
an estimate of the most probable companion mass.  For J0112+1835, the most likely
companion is a 0.85 \msun\ WD at an orbital separation of 1.2 \rsun .

	There is no evidence for a 0.85 \msun\ WD in the spectrum of J0112+1835, nor
do we except there to be.  If we
pessimistically assume that the two WDs formed at the same time 100 Myr - 1 Gyr ago,
we would expect the 0.85 \msun\ companion to have $M_g=11-13$ mag \citep{bergeron95};
it would be 15 - 100 times less luminous than the
0.16 \msun\ WD. Of course to form a short-period ELM WD binary like J0112+1835
requires two consecutive phases of common-envelope evolution in which the ELM WD is
created last, giving the more massive secondary yet more time to cool and fade.

	To understand the evolutionary history of J0112+1835 and our other ELM WD
binaries requires assumptions about the energy balance and angular momentum balance
of the common envelope phase \citep[e.g.][]{nelemans05b}. \citet{kilic10} discuss
one possible origin for a $\sim1$ h orbital period ELM WD evolving from a system
containing a 3 \msun\ and a 1 \msun\ star.  The 3 \msun\ star evolves off the main
sequence, overflows its Roche lobe as a giant with a 0.6 \msun\ core, forms a helium
star (sdB) which does not expand after He-exhaustion in the core, and turns into a
WD.  The 1 \msun\ star also overflows its Roche lobe after main-sequence evolution
when its core is around 0.2 \msun .  We can estimate orbital separations if we
assume that the evolved stars exactly fill their Roche lobes.  In this case, the
first common-envelope phase has an orbital separation of 860 \rsun\ and the second
common-envelope phase has an orbital separation of 25 \rsun .  The orbital
separation of J0112+1835 and our other ELM WD binaries is now $\simeq$1 \rsun .

	General relativity predicts that short period binaries like J0112+1835 lose
energy and angular momentum to gravitational wave radiation.  The time scale for the
binary to shrink and begin mass transfer via Roche-lobe overflow is given by the
gravitational wave merger time
	\begin{equation}
\tau = \frac{(M_1 + M_2)^{1/3}}{M_1 M_2} P^{8/3} \times 10^{-2} {\rm Gyr}
	\end{equation} where the masses are in \msun\ and the period $P$ is in hours
\citep{landau58}.  Inserting the minimum companion mass yields the maximum merger
time given in Table \ref{tab:orbit}.  For the most probable companion mass of 0.85
\msun , J0112+1835 will begin mass transfer in 2.1 Gyr.  \citet{kilic10} discuss the
many possible stellar evolution paths for such a system.  This system's mass ratio
$M_1/M_2\leq$0.26 suggests that mass transfer will be stable \citep{marsh04} and
that J0112+1835 will likely evolve into an AM CVn system.

\subsection{J015213.77+074913.9}

	The ELM WD J0152+0749 has a longer orbital period of $7.749\pm0.003$ hr and
a radial velocity amplitude of $434\pm4$ \kms .  There is a 74\% probability that
the companion is a WD with $<$1.4 \msun\ and a 13\% probability that the companion
is a neutron star with 1.4-3.0 \msun .  For $i=60\arcdeg$, the most likely companion
is a 0.78 \msun\ WD at an orbital separation of 1.9 \rsun .  This system will not
begin mass transfer within a Hubble time.

\subsection{J144342.74+150938.6}

	The ELM WD J1443+1509 has a best-fit orbital period of 4.573 hr.  However,
the current data set (which spans only 3 nights) allows for a significant alias at
5.75 hr.  The relatively large $614\pm6$ \kms\ radial velocity amplitude of this
system implies that the companion must be relatively massive, regardless of the
exact period.  Adopting the best-fit orbital period, there is a 60\% probability
that the companion is a WD with $<$1.4 \msun\ and a 20\% probability that the
companion is a neutron star with 1.4-3.0 \msun .  For $i=60\arcdeg$, the most likely
companion is a 1.15 \msun\ WD at an orbital separation of 1.5 \rsun .

	This system will begin mass transfer in less than 4.1 Gyr.  The likelihood
that the system contains a pair of WDs whose total mass exceeds the Chandrasekhar
mass is 6\%. Given the observed mass ratio $M_1/M_2\leq$0.20, this system will
undergo stable mass transfer and will likely evolve into an AM CVn system.

\subsection{J151826.68+065813.2}

	The ELM WD J1518+0658 is similar to J0152+0749 except that the secondary is
likely at a larger orbital separation.  The system has an orbital period of
$14.624\pm0.001$ hr and a radial velocity amplitude of $344\pm4$ \kms .  Given these
parameters, there is a 74\% probability that the companion is a WD with $<$1.4
\msun\ and a 13\% probability that the companion is a neutron star with 1.4-3.0
\msun .  For $i=60\arcdeg$, the most likely companion is a 0.78 \msun\ WD at an
orbital separation of 3.0 \rsun .  This system will not begin mass transfer within a
Hubble time.

\subsection{J174140.49+652638.7}

	The ELM WD J1741+6526 is arguably the most interesting of the six new
systems.  J1714+6526 has an orbital period of $1.4666\pm0.0001$ hr and a radial
velocity amplitude of $986\pm4$ \kms .  Because our 6 minute exposure times span 7\%
of its orbital phase ($\delta \phi = 0.43$ radians), the observed amplitude is
underestimated by a factor of $\sin{\delta \phi}/\delta \phi=0.97$.  The true radial
velocity amplitude of the ELM WD is thus 1,016 \kms .

	Using the corrected orbital parameters, there is a 43\% probability that
J1741+6526's companion is a WD with $<$1.4 \msun\ and a 31\% probability that the
companion is a neutron star with 1.4-3.0 \msun .  For $i=60\arcdeg$, the most likely
companion is a 1.55 \msun\ neutron star at an orbital separation of 0.8 \rsun .  
Given that this putative neutron star must have accreted material from the common
envelope evolution of the ELM WD progenitor, a neutron star companion is possibly a
milli-second pulsar.

	Milli-second pulsars in short-period orbits are difficult to detect because
of their rapidly changing velocity, but are valuable probes of general relativity
and gravitational wave physics.  For the most likely companion mass of 1.55 \msun ,
J1741+6526 will merge in 130 Myr -- twice as fast as the Hulse-Taylor pulsar.  The
gravitational wave strain for this system $h\simeq7\times10^{-23}$ is in principle
detectable by the proposed $LISA$ mission, but its orbital frequency places the
system below the expected confusion-limit from other double-degenerate gravitational
wave sources \citep{roelofs07a}.  We are pursuing follow-up radio and X-ray
observations to test whether or not J1741+6526 is a milli-second pulsar binary.

	If the companion is a massive WD, on the other hand, there is a 30\%
likelihood that J1741+6526 contains a pair of WDs whose total mass exceeds the
Chandrasekhar mass.  With a $M_1/M_2\leq$0.15 mass ratio, J1741+6526 will initially
evolve into a stable mass-transfer AM CVn system.  When the massive WD accretes
sufficient mass, it is possible J1741+6526 will explode as a Type Ia supernova.  If
the companion is not massive enough to be a Ia progenitor, the system will likely
create an underluminous .Ia explosion \citep{bildsten07}. We are pursuing time
series photometry of J1741+6526 (Hermes et al., in prep.) to better constrain the
orbital inclination and future evolution of this system.

\subsection{J184037.78+642312.3}

	The ELM WD J1840+6423 has a $4.5912\pm0.0012$ hr orbital period, with
a significant alias at 3.85 hr, and a $544\pm4$ \kms\ radial velocity amplitude.  
Assuming the best-fit orbital period, there is a 70\% probability that the companion
is a WD with $<$1.4 \msun\ and a 15\% probability that the companion is a neutron
star with 1.4-3.0 \msun .  For $i=60\arcdeg$, the most likely companion is a 0.88
\msun\ WD at an orbital separation of 1.4 \rsun .

	This system is similar to J1443+1509:  J1840+6423 will begin mass transfer
in less than 5.0 Gyr, and the likelihood that the system contains a pair of WDs
whose total mass exceeds the Chandrasekhar mass is 4\%.  Given the observed mass
ratio $M_1/M_2\leq$0.27, this system should undergo stable mass transfer and will
likely evolve into an AM CVn system.

\section{DISCUSSION}

	The ELM Survey has now identified 19 merging ELM WD systems that will
coalesce in $<$10 Gyr.  The first 12 were summarized by \citet{kilic11a}.  Three
more systems were published earlier this year:  two 39 minute orbital period
binaries \citep{kilic11b, kilic11c} and one 12 minute period eclipsing binary
\citep{brown11b}.  In this paper we add four more merging systems to the count.  To
aid the reader, Table \ref{tab:summary} summarizes the properties of the 19 merging
ELM WD systems.

\begin{figure}		% FIGURE 7:  MERGE PLOT
 % \plotone{/pool/wbrown0/Bcand/LMWD3/SPY/elm3.ps}
 \plotone{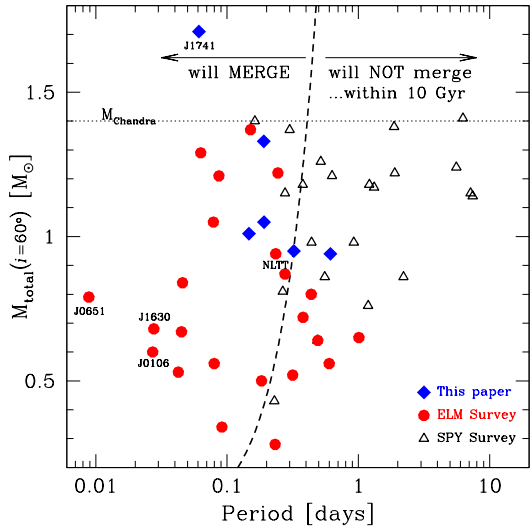}
 \caption{ \label{fig:merge}
	Binary orbital period versus total system mass for the full ELM Survey and
for the Supernova Progenitor Survey \citep[SPY,][]{koester09}.  We plot total system 
mass assuming $i=60\arcdeg$ when orbital inclination is unknown, and the correct
system mass when inclination is known.
        WD binaries from SPY are drawn with open triangles, previously published ELM
Survey binaries are drawn with solid red circles, and the 6 new ELM WD binaries from
this paper are drawn with solid blue diamonds.  The dashed line shows the threshold
at which a 1:1 mass ratio system will merge in 10 Gyr.}
 \end{figure}

%TABLE4
\begin{deluxetable*}{ccccccccrccl}
\tabletypesize{\scriptsize}
\tablecolumns{12}
\tablewidth{0pt}
\tablecaption{Merger Systems in the ELM Survey\label{tab:summary}}
\tablehead{
	\colhead{Object}&
	\colhead{$T_{\rm eff}$}&
	\colhead{$\log g$}&
	\colhead{Mass}&
	\colhead{$P$}&
	\colhead{$K$}&
	\colhead{$M_2$}&
	\colhead{$\tau_{\rm merge}$}&
	\colhead{$\gamma$}&
	\colhead{$\mu_{\rm RA}$} &
	\colhead{$\mu_{\rm Dec}$}&
	\colhead{Ref}\\
  & (K) &  & \msun  & days & \kms & \msun  & Gyr & \kms & mas yr$^{-1}$ & mas yr$^{-1}$ &
}
	\startdata
J0022$-$1014 & 18980 & 7.15 & 0.33 & 0.07989 & 145.6 & $\geq 0.19$ & $\leq 0.73$ & $  81$ & $ -7.8$ & $-13.2$ & 5 \\
J0106$-$1000 & 16490 & 6.01 & 0.17 & 0.02715 & 395.2 &       0.43  &      0.037  & $   2$ & $ 20.2$ & $ -9.6$ & 6 \\
J0112+1835   &  9770 & 5.57 & 0.16 & 0.14698 & 295.3 & $\geq 0.62$ & $\leq 2.67$ & $-121$ & $  7.8$ & $-17.4$ & 0 \\
J0651+2844   & 16400 & 6.79 & 0.25 & 0.00885 & 657.3 &       0.55  &     0.0009  & $  17$ & $ -3.6$ & $ -1.2$ & 2 \\
J0755+4906   & 13160 & 5.84 & 0.17 & 0.06302 & 438.0 & $\geq 0.81$ & $\leq 0.22$ & $ -51$ & \nodata & \nodata & 1 \\
J0818+3536   & 10620 & 5.69 & 0.17 & 0.18315 & 170.0 & $\geq 0.26$ & $\leq 8.89$ & $-201$ & \nodata & \nodata & 1 \\
J0822+2753   &  8880 & 6.44 & 0.17 & 0.24400 & 271.1 & $\geq 0.76$ & $\leq 8.42$ & $ -52$ & $  3.4$ & $-19.2$ & 3 \\
J0849+0445   & 10290 & 6.23 & 0.17 & 0.07870 & 366.9 & $\geq 0.64$ & $\leq 0.47$ & $  48$ & $ -0.7$ & $ -0.4$ & 3 \\
J0923+3028   & 18350 & 6.63 & 0.23 & 0.04495 & 296.0 & $\geq 0.34$ & $\leq 0.13$ & $   2$ & $ -4.2$ & $-24.9$ & 1 \\
J1053+5200   & 15180 & 6.55 & 0.20 & 0.04256 & 264.0 & $\geq 0.26$ & $\leq 0.16$ & $  12$ & $-29.7$ & $-31.2$ & 3,8 \\
J1233+1602   & 10920 & 5.12 & 0.17 & 0.15090 & 336.0 & $\geq 0.86$ & $\leq 2.14$ & $ -35$ & \nodata & \nodata & 1 \\
J1234$-$0228 & 18000 & 6.64 & 0.23 & 0.09143 &  94.0 & $\geq 0.09$ & $\leq 2.69$ & $  94$ & $-14.5$ & $-12.3$ & 5 \\
J1436+5010   & 16550 & 6.69 & 0.24 & 0.04580 & 347.4 & $\geq 0.46$ & $\leq 0.10$ & $ -30$ & $  7.8$ & $ -5.1$ & 3,8 \\
J1443+1509   & 14770 & 6.06 & 0.17 & 0.19053 & 306.7 & $\geq 0.83$ & $\leq 4.09$ & $-172$ & $-30.9$ & $-53.5$ & 0 \\
J1630+4233   & 14670 & 7.05 & 0.30 & 0.02766 & 295.9 & $\geq 0.30$ & $\leq 0.03$ & $ -11$ & $  2.3$ & $ -7.3$ & 7 \\
J1741+6526   &  9900 & 5.20 & 0.16 & 0.06111 & 508.0 & $\geq 1.09$ & $\leq 0.17$ & $ -70$ & $ -3.2$ & $ -4.6$ & 0 \\
J1840+6423   &  9100 & 6.22 & 0.17 & 0.19130 & 272.0 & $\geq 0.64$ & $\leq 5.00$ & $ -76$ & $-12.4$ & $-26.7$ & 0 \\
J2119$-$0018 & 10360 & 5.36 & 0.17 & 0.08677 & 383.0 & $\geq 0.75$ & $\leq 0.54$ & $ -28$ & \nodata & \nodata & 1 \\
NLTT 11748   &  8690 & 6.54 & 0.18 & 0.23503 & 273.4 &       0.76  &       7.20  & $ 126$ & $236.1$ & $-179.2$ & 4,9,10 \\
	\enddata
\tablerefs{ (0) this paper; (1) \citet{brown10c}; (2) \citet{brown11b}; (3) \citet{kilic10}; 
	(4) \citet{kilic10b}; (5) \citet{kilic11a}; (6) \citet{kilic11b}; (7) \citet{kilic11c}; 
	(8) \citet{mullally09}; (9) \citet{steinfadt10}; (10) \citet{kawka10} }

\tablecomments{ Measurement errors reported in the references.  Proper motions
are from USNO-B+SDSS \citep{munn04}. } 

\end{deluxetable*}

	While the merging ELM WD sample is not complete, properties such as orbital
period and secondary mass are in principle independent of color and magnitude.  
Thus we split the merging ELM WD sample into thirds and search for significant
correlations among the sample properties.

	Looking at Table \ref{tab:summary}, the six ELM WDs with $<$1.1 hr orbital
periods are 6000$\pm$2400 K hotter than the seven ELM WDs with $>$3.5 hr orbital
periods.  In other words, we observe an absence of cool ELM WDs with short orbital
periods.  This period-temperature dependence makes sense if the ELM WDs in short
orbital period systems merge before they cool.\footnote{The period-temperature
effect is unchanged if we drop the 12 minute orbital period system, which is
possibly heating up due to tidal effects \citep{piro11, fuller11}.} We take this
correlation as evidence that common envelope evolution creates ELM WDs directly in
short orbital period systems.

	The minimum companion masses of the ELM WDs span an order of magnitude, from
0.1 \msun\ to 1.1 \msun , and also appear to correlate with ELM WDs' temperatures.  
The coolest seven ELM WDs, those with $T_{\rm eff}$$<$10,500 K, have minimum
companion masses that are 0.41$\pm$0.23 \msun\ larger than the hottest six ELM WDs
with $T_{\rm eff}$$>$16,000 K.  This result largely comes from our present survey,
which targets relatively cool ELM WDs and finds only extreme mass-ratio binaries.
Our interpretation is that shorter period binary systems experience increased mass
loss during their evolution and end up with lower mass companions.  A larger sample
is required to establish these trends with increased significance.

	Figure \ref{fig:merge} compares the distribution of binary orbital period
and total system mass for the ELM Survey with that of the Supernova Progenitor
Survey \citep[SPY,][]{napiwotzki01, koester09}.  The SPY survey has measured the
velocity variability of $\simeq$1000 WDs, the largest survey of its kind.  Our
Figure is an adaptation of \citet{geier10}'s Figure 4, where we plot total system
mass assuming $i=60\arcdeg$ when orbital inclination is unknown, and the correct
system mass when inclination is known via ellipsoidal variations and/or eclipses. It
is notable that the SPY survey, which samples the full WD population, finds only a
handful of systems with orbital periods short enough that they might possibly merge
in less than 10 Gyr.

	The ELM survey, which targets $\sim$0.2 \msun\ WDs, finds almost every
object in a system with $\lesssim$1 day orbital period, the majority of which will
merge due to gravitational wave radiation in less than 10 Gyr.  The merging ELM WD
systems are unlikely Type Ia supernovae progenitors, however, because the total mass
of the systems is likely below the Chandrasekhar mass.  Their most likely
evolutionary futures include the formation of stable mass-transfer AM CVn systems
and underluminous supernovae, or unstable mass-transfer mergers that form extreme
helium stars (RCrB) and single helium-enriched subdwarf O stars \citep[discussed
further in][]{kilic10}. One approach to constrain these scenarios is to compare ELM
WD merger rates with the formation rates of different classes of objects as
attempted in \citet{brown11a}.  A larger, well-defined sample will provide improved
constraints, as will follow-up light curves that directly measure the orbital
inclination and nature of the unseen binary companions.

	We close by noting that at least one-third of the merging ELM WD systems
have the kinematics and locations of halo objects.  Looking at Table
\ref{tab:summary}, four ELM WD systems (J0112, J0818, J1443, NLTT 11748) have
systemic radial velocities $|\gamma|> 100$ \kms .  Proper motions with 5 mas
yr$^{-1}$ uncertainties are available for 15 of the 19 objects \citep{munn04,
lepine05}.  Combining radial velocities and proper motions reveals five systems
(J0106, J0818, J1053, J1443, NLTT 11748) with total space velocities $>$200 \kms\
with respect to the Sun.  Thus 6 unique ELM WD systems (32\%) have kinematics that
indicate a halo origin.  In addition to motions, physical locations also support a
halo origin.  The ELM WDs are not clustered at low Galactic latitudes in our survey,
as expected for a disk population, but rather are found equally at high and low
Galactic latitudes.  The typical ELM WD in our survey has median luminosity
$M_g=8.5$, apparent magnitude $g_0=18.8$, and Galactic latitude $b=43\arcdeg$, and
thus is located $\simeq$1 kpc above the Galactic plane.  We conclude that ELM WD
systems continue to form and merge in both the disk and the halo, a conclusion that
has implications for gravitational wave source predictions \citep{ruiter09}.

\section{CONCLUSION}

	We present a targeted spectroscopic survey of ELM WDs candidates selected by
color.  The survey is successful:  it is now 71\% complete and has uncovered 18 new
ELM WDs.  Of the 7 ELM WDs with follow-up observations, 6 are compact binary systems
and 4 have gravitational wave merger times less than 5 Gyr.  The most intriguing new
object is J1741+6526, which likely has either a milli-second pulsar binary companion
or a massive WD companion making the system a possible supernova Type Ia or .Ia
progenitor.  Follow-up observations are underway to establish the nature of this
system as well as the other ELM WDs.

	Based on these initial results, we expect that completing our targeted ELM
WD survey will double the number of merging systems in our sample from 19 to
$\simeq$40 systems.  We expect that photometric follow-up will reveal additional
eclipsing systems.
	Our efficiency for ELM WDs discoveries increases with apparent magnitude
such that, if we were to expand our discovery survey to $g\simeq20.5$, we could in
principle double again our sample of ELM WDs (although the observations are more
expensive).  The absence of short-merger time systems at cooler temperatures
suggests that expanding our survey to hotter objects may yield additional $\sim$10
min orbital period ($<$1 Myr merger time) systems.  These are directions we will
pursue in upcoming observing runs. With a sample of $\sim$100 ELM WDs systems
spanning a well-defined range of temperature, we look forward to placing robust
constraints on the role of these detached double degenerate binaries as supernovae
progenitors and gravitational wave sources.

\acknowledgments
	We thank M.\ Alegria, J.\ McAfee, and A.\ Milone for their assistance with
observations obtained at the MMT Observatory, and P.\ Berlind and M.\ Calkins for
their assistance with observations obtained at the Fred Lawrence Whipple
Observatory.  We thank A.\ Piro for a discussion about hot ELM WDs. This project
makes use of data products from the Sloan Digital Sky Survey, which is managed by
the Astrophysical Research Consortium for the Participating Institutions.  This
research makes use of NASA's Astrophysics Data System Bibliographic Services.  This
work was supported in part by the Smithsonian Institution.  MK was supported in part
by NASA through the {\em Spitzer Space Telescope} Fellowship Program, under an award
from CalTech.

{\it Facilities:} \facility{MMT (Blue Channel Spectrograph)}

% \clearpage
	% REFERENCES
% \bibliographystyle{/home/wbrown/lib/apj} \bibliography{/home/wbrown/text/RefHS}

\appendix \section{DATA TABLE}

	Table 4 presents our radial velocity measurements. The Table columns include 
object name, heliocentric Julian date (based on UTC), heliocentric radial velocity 
(uncorrected for the WD gravitational redshift), and velocity error.
	~~~~ [See on-line journal or tab4.dat in the arXiv submission.]

\end{document}